\documentclass[prc,twocolumn,showpacs,superscriptaddress,preprintnumbers,amsmath,amssymb]{revtex4}

\usepackage[dvipdfmx]{graphicx}
\usepackage{dcolumn}
\usepackage{bm}
\usepackage{amsmath}

\newcommand{\Slash}[1]{{\ooalign{\hfil/\hfil\crcr$#1$}}}
  
\usepackage[normalem]{ulem}  
\usepackage[dvips]{color} 

\renewcommand\sout{\bgroup \color{red} \ULdepth=-.5ex \ULset}

\begin{document}
\preprint{RIKEN-QHP-116}
\title{Momentum transport away from a jet in an expanding nuclear medium}

\author{Yasuki Tachibana}
\email{tachibana@nt.phys.s.u-tokyo.ac.jp}

\affiliation{Department of Physics, The University of Tokyo, Tokyo 113-0033, Japan}
\affiliation{%
Theoretical Research Division, Nishina Center, RIKEN, Wako 351-0198, Japan
}
\affiliation{%
Department of Physics, Sophia University, Tokyo 102-8554, Japan
}

\author{Tetsufumi Hirano}
\email{hirano@sophia.ac.jp}
\affiliation{%
Department of Physics, Sophia University, Tokyo 102-8554, Japan
}

\date{\today}

\begin{abstract}
We study 
the transport dynamics of 
momenta deposited from jets
in ultrarelativistic heavy-ion collisions.
Assuming that the high-energy partons
traverse expanding quark-gluon fluids
and are subject to lose their energy and momentum,
we simulate dijet asymmetric events
by solving relativistic hydrodynamic equations 
numerically without linearization in the fully (3+1)-dimensional coordinate. 
Mach cones are formed and  
strongly broadened by radial flow of the background medium. 
As a result, 
the yield of low-$p_{T}$ particles increases
at large angles from the jet axis 
and compensates the dijet momentum imbalance
inside the jet-cone. 
This provides 
an intimate link 
between 
the medium excitation by jets 
and 
results 
in dijet asymmetric events 
observed by the CMS Collaboration. 
\end{abstract}
\maketitle

\emph{Introduction.}---
The quark gluon plasma ({QGP}),
supposed to have filled 
the early universe a few microseconds after the Big Bang, 
is 
the deconfined state of quarks and gluons 
realized under an extremely hot and dense
condition \cite{Yagi:2005yb}. 
In heavy ion collisions 
at the Relativistic Heavy Ion Collider ({RHIC}) at BNL  
and the Large Hadron Collider ({LHC}) at {CERN}, 
the {QGP} is created experimentally. 
By colliding relativistically accelerated heavy nuclei, 
extremely high-temperature is achieved 
in the experiments. 
From the analysis of 
experimental data of elliptic flow, 
it has turned out that 
the {QGP} behaves like an almost-perfect fluid 
because of the strong interacton 
among the constituent particles 
\cite{Heinz:2001xi, sQGP1, sQGP2, sQGP3, Hirano:2005wx}. 

Jets, namely partons with large transverse momenta, are created
in hadron or nuclear collisions at collider energies.
In nuclear collisions, these  partons are subject to traverse a hot and dense medium.
While traversing the medium, the parton loses its energy 
through strong interaction between them 
\cite{Bjorken:1982tu, Appel:1985dq, Blaizot:1986ma, Rammerstorfer:1990js, Gyulassy:1990ye, Gyulassy:1993hr, Thoma:1990fm}. 
Through the amount of lost energy,
one can extract one of the fundamental properties of the medium, 
namely stopping power of the QGP against
high-energy partons. 
In addition, 
the energy-momentum deposition from jets 
excites the medium and  
propagation of this medium excitation 
may give information about 
the transport coefficients and the sound velocity of the QGP. 
Thus jet quenching phenomena provide 
a unique opportunity 
to probe the properties of 
the primordial matter composed of elementary particles 
in quantum chromodynamics (QCD). 

The next question is 
where and how 
this lost energy diffuses inside the medium. 
In experiments, 
a large number of low-$p_{T}$ hadrons 
at large angles from an axis of the quenched jet 
is observed in Pb-Pb collisions at the {LHC} \cite{Chatrchyan:2011sx}. 
The total transverse momentum of 
these low-$p_{T}$ particles 
together with the quenched jet 
balances 
that of a jet propagating 
in the opposite direction. 
Therefore, 
the lost energy of the quenched jet 
can be considered as a source of
the low-$p_{T}$ particles emitted at large angles from the jet axis. 
On the theoretical side, 
the conical flow, the so-called Mach cone, is supposed to develop  
in the QGP 
as interference of sound waves 
induced by the supersonic jet particles 
\cite{Stoecker:2004qu, CasalderreySolana:2004qm}.
This conical flow propagates 
at a specific angle 
from the direction of the energetic partons. 
The Mach cone has been analyzed 
in various theoretical studies such as
hydrodynamics with 
\cite{CasalderreySolana:2004qm, Neufeld:2008fi, Neufeld:2009ep, Qin:2009uh, Neufeld:2010tz}
or without 
\cite{oai:arXiv.org:nucl-th/0503028, Betz:2010qh, Tachibana:2012sa} 
linearlization,
{AdS/CFT} calculations 
\cite{Gubser:2007ga, Chesler:2007sv, Noronha:2008un}, 
and a parton-transport model  
\cite{Bouras:2012mh, Bouras:2014rea}.
Here we emphasize that
the background QGP medium is no longer static, but expands
with relativistic flow velocity.
The resultant Mach cones should be distorted by the expansion of the QGP 
\cite{Satarov:2005mv, oai:arXiv.org:nucl-th/0503028, Betz:2010qh, Tachibana:2012sa, Bouras:2014rea}.

In this Rapid Communication, 
the dynamical transport process 
of energy and momentum 
deposited from energetic partons traversing 
the expanding QGP fluid is studied. 
We show that 
the low $p_T$ enhancement 
at large angles from the jet axis 
as observed at the LHC 
is a consequence of energy-momentum deposition of 
high-energy partons and its 
transport in the medium.
Here we use relativistic hydrodynamic 
framework 
to describe the medium response.
This is the first attempt to numerically solve 
relativistic hydrodynamic equations with source terms 
without linearization 
in fully (3+1)-dimensional Milne coordinates.
In this way, 
we properly consider the interplay dynamics 
between the hydrodynamical expansion of the QGP 
and the collective flow 
induced by the energy-momentum deposition of jets. 

In the following, 
we first overview
the current experimental situation of dijet asymmetry
in high energy nuclear collisions at the LHC energy.
Motivated by these findings, 
we formulate relativistic hydrodynamic equations 
with source terms 
which correspond to deposition of jets' energies and momenta.  
We next solve these fully nonlinear hydrodynamic equations
without resorting to linearization of the equations.
Finally, we investigate energy and momentum balance in dijet events. 

\emph{Dijet asymmetry.}---
At the leading order, 
back-to-back partons are created 
with equal transverse momenta. 
Compared with the dijet events 
in hadron-hadron collisions, 
jet energies are more imbalanced 
due to jet quenching in nuclear collisions: 
One parton going toward the outside of the medium is observed as a leading jet 
and the other one going inside is observed as a subleading jet. 
Thus the amount of the lost energy is different between the pair
due to the position of the pair creation.
The asymmetry ratio to quantify 
the dijet transverse momentum imbalance  
is defined as
\begin{eqnarray}
A_{J} = \frac{p_{T,1}-p_{T,2}}{p_{T,1}+p_{T,2}},
\end{eqnarray}
where $p_{T,1}$ and $p_{T,2}$ 
are the transverse momentum of the leading jet and of the subleading jet, respectively. 
In central Pb-Pb collisions at LHC, 
a mean shift of $A_{J}$ to higher values is observed \cite{Aad:2010bu, Chatrchyan:2011sx}. 
The increase of highly asymmetric dijet events
compared with $p+p$ collisions 
indicates a substantial amount of jet energy loss 
at the {LHC}. 
The dijet asymmetry at the {LHC} 
has been explained 
theoretically 
from medium modification of jet in the {QGP}
\cite{CasalderreySolana:2010eh, Qin:2010mn, Young:2011qx, He:2011pd}. 
To see the balance of the whole transverse momentum in an event, 
one can define transverse momentum along the jet axis as
\begin{eqnarray}
\Slash{p}^{||}_{T} = \sum_{i}-p^{i}_{T}\cos(\phi_{i}-\phi_{1}),
\end{eqnarray}
where the sum is taken over all tracks in a dijet event 
and its transverse momentum is projected 
onto the subleading jet axis $\phi_2 = \phi_1 + \pi$ in the azimuthal direction.
$\Slash{p}^{||}_{ T}$  
is measured in dijet events 
in Pb-Pb collisions at the LHC by the {CMS} Collaboration \cite{Chatrchyan:2011sx}. 
The transverse momentum averaged over events, $\langle \Slash{p}^{||}_{ T} \rangle$,
turned out to vanish within uncertainties even in large dijet asymmetric events. 
The leading jet dominantly contributes 
to negative $\langle \Slash{p}^{||}_{T} \rangle$, 
which is balanced by
the lower momentum particles with
$0.5< p_{T} < 8$ GeV/$c$ in the direction of  the subleading jet
outside the cone 
$\Delta R = \left[(\Delta \phi)^2 + (\Delta \eta)^2)\right]^{1/2}>0.8$.
Thus an apparent imbalance of the dijet momenta 
only inside the cone
is compensated by the low-$p_{T}$ particles 
at large angles 
from the jet axis.
Since the low-$p_{T}$
particles play an important role in momentum balance of dijet asymmetric events,
it has been suggested that the energy deposition 
from the traveling partons wakes the {QGP} medium 
and induces collective flow to
enhance
low momentum particles 
at large angles from the axis of the quenched jet.

\emph{Hydrodynamic model with source terms.}---
Motivated by these observations,  
we investigate the mechanism of
energy and momentum 
transported away from the jet 
in dijet events.
Assuming local thermal equilibrium, 
we perform relativistic hydrodynamic simulations
to describe 
the spacetime evolution of the {QGP} medium. 
We introduce source terms in the hydrodynamic equations 
which exhibit the energy and momentum deposition from these partons: 
\begin{eqnarray}
\partial_{\mu}T^{\mu \nu}\left(x\right)=J^{\nu}\left(x\right). \label{eqn:hydro_source}
\end{eqnarray}
Here, $T^{\mu \nu}$ is the energy-momentum tensor of the {QGP} fluid 
and 
$J^{\nu}$ is the four-momentum density   
deposited from the traversing jet partons. 
We solve 
the nonlinear hydrodynamic equations 
numerically 
without linearization 
with a new high-precision scheme 
in fully (3+1)-dimensional 
coordinates \cite{Murase:prep}. 

For perfect fluids, the energy momentum tensor can be decomposed as 
\begin{eqnarray}
T^{\mu \nu}&=&(e+P)u^{\mu}u^{\nu}-Pg^{\mu \nu},
\end{eqnarray}
where $e,\:P$, $\:u^{\mu}$, 
and $g^{\mu\nu}={\rm diag}\left(1, -1, -1, -1\right)$ 
are energy density, pressure, four-flow velocity, 
and the Minkowski metric, respectively. 
Assuming that the lost energy and momentum 
of energetic partons are 
instantaneously 
deposited and thermalized 
in a {QGP}-fluid cell,  
we employ a simple form of the source terms 
for a pair of massless particles traveling through the medium 
\begin{eqnarray}
J^{\mu}\left(x\right)&=&\sum_{a=1,2}J_{a}^{\mu}\left(x\right),\label{eqn:source_sum}\\ 
J_{a}^{0}\left(x\right)&=&
-\frac{dp_{a}^0}{dt} 
\delta^{(3)}\left(\mbox{\boldmath $x$}-\mbox{\boldmath $x$}_{a}(t)\right),\label{eqn:source_tm}\\ 
\mbox{\boldmath$J$}_{a}(x)&=&\frac{\mbox{\boldmath$p$}_{a}}{p_{a}^0}J_{a}^{0}(x), \label{eqn:source_sp}
\end{eqnarray}
where the index $a$ denotes each energetic parton 
which is to be observed as the leading ($a=1$) or subleading jet ($a=2$). 
For a given equation of state, we solve Eq.~(\ref{eqn:hydro_source}) 
in the $\left(3+1\right)$-dimensional Milne coordinates 
$\left(\tau, x, y, \eta_{s}\right)$ 
numerically without linearization. 
$\tau=\left(t^2+z^2\right)^{1/2}$ is the proper time 
and $\eta_{s}=\left(1/2\right)\ln\left[\left(t+z\right)/\left(t-z\right)\right]$ is 
the spacetime rapidity. 
As an equation of state, 
we employ that of the ideal gas of massless quarks and gluons, 
$P(e)=e/3$, for simplicity.
In this framework, 
we can handle an expanding background QGP fluid created in heavy ion collisions
together with its response to propagation of dijets. 
Very small deposited energy and momentum 
relative to the total energy and momentum of the medium 
are treated here, 
so it is necessary to keep the energy-momentum conservaton 
in the whole system at a very high precision. 
A new and robust scheme, which we developed and employed here, 
plays a crucial role to conserve the energy and momentum 
of the fluid accurately 
in full $\left(3+1\right)$-dimensional Milne coordinates 
and is essential for this calculation \cite{Murase:prep}.

We set up the initial QGP fluid at $\tau_0 = 0.6\:{\rm fm}/c$. 
Around the mid-rapidity region, the initial energy density 
is flat in the $\eta_{s}$ direction like the Bjorken scaling solution \cite{Bjorken:1982qr}.   
The flat region is smoothly connected to vacuum
at the both ends by using  a half Gaussian \cite{Hirano:2001eu}:   
\begin{eqnarray}
H\left(\eta_{s} \right)\!&=&\!
\exp\!\left[-\frac{\left(\left|\eta_{s}\right|-\eta_{\rm flat}/2\right)^2} {2\sigma_{\eta}^2} \theta \left(\left|\eta_{ s}\right|-\frac{\eta_{\rm flat}}{2} \right) \right], 
\end{eqnarray}
where $\eta_{\rm flat}$ and $\sigma_{\eta}$ are the rapidity length of the flat region and the width of the Gaussians, respectively.
Then full initial energy density distribution is factorized as 
\begin{eqnarray}
e\left(\tau=\tau_{0},\:x,\:y,\:\eta_{\rm s}\right)=e_{T}\left(x,y\right)H\left(\eta_{\rm s} \right).  
\end{eqnarray}
Here $e_{T}$ is the smooth transverse profile of the initial energy density 
for central ($0$-$5\%$) Pb-Pb collisions. 
We calculate 
the number density of participants and binary collisions by using 
Monte Carlo Glauber model. We assume that the entropy density distribution is
proportional to the linear combination of these two densities.
The distribution is normalized by comparison of final multiplicity at mid-rapidity with the LHC data \cite{Hirano:2010je}. 
Then, by using the equation of state, $e_{T}$ is obtained. 
We choose $\eta_{\rm flat}=10$ and $\sigma_{\eta}=0.5$ 
for Pb-Pb collision at the {LHC} \cite{Schenke:2011tv}. 
For the energy loss of the partons 
in the local rest frame of the fluid, 
we employ the collisional energy loss \cite{Thomas:1991ea} 
\begin{eqnarray}
-\frac{dp_a^0}{dt} = A\times \frac{8}{3}\pi{\alpha_{ s}}^2T^2\left(1+\frac{1}{6}n_{ f}\right)\log\frac{\sqrt{4Tp_{a}^0}}{m_{D}}. \label{eqn:source}
\end{eqnarray}
Here, $p_{a}^0$ is the energy of a jet particle in the local rest frame, 
$\alpha_{s}=g^2/\left(4\pi\right)$ is the strong coupling constant, 
$n_{f}$ is the number of active flavors in the QGP medium, 
$m_{D}=\left(1+\frac{1}{6}n_{f}\right)^{1/2}gT$ is the Debye mass, 
and $A$ is a parameter which allows us to control the strength of the energy loss. 
Here, we set $n_{f}=3$ (u, d, s), $\alpha_{s}=0.3$, and $A=15$. 
The source terms (\ref{eqn:source_sum}) are obtained from 
Eqs. (\ref{eqn:source_tm}), (\ref{eqn:source_sp}), and (\ref{eqn:source}), 
then Lorentz boosted to the Milne coordinates. 

The Cooper-Frye formula \cite{Cooper:1974mv} is used 
 to obtain the momentum distribution of particle species $i$ 
from hydrodynamic outputs,
\begin{equation}
p^0 \frac{dN_i}{d^3p}=
\frac{g_i}{\left(2\pi\right)^3}\int \frac{p^{\mu}d\sigma_{\mu}\left(x\right)}{\exp\left[p^{\mu}u_{\mu}\left(x\right)/T\left(x\right) \right]\pm 1}, 
\label{eqn:Cooper-Frye}
\end{equation}
where $g_i$ is the degeneracy and $\pm$ corresponds to Fermi or Bose distribution for particle species $i$. 
The freeze-out is supposed to occur 
at fixed proper time $\tau_{f}=9.6\:{\rm fm}/c$, 
which is a typical value for central Pb-Pb collisions 
and 
not crucial for results presented here. 
Thus $p^{\mu} d\sigma_{\mu}=p_{T}\cosh\left(\eta_{p}-\eta_{s}\right)\tau_{f}dxdyd\eta_{s}$,
where $\eta_{p}$ is the momentum rapidity. 
Here we set as a rapidity cut $|\eta_{p}|<2.4$.
Suppose $\langle \Slash{p}^{||}_{T} \rangle = \langle - p_{T} \cos \left(\phi_{p}-\phi_{1}\right)\rangle$ 
contains 
contribution from particles originated from fluids,
it is calculated from Eq.~(\ref{eqn:Cooper-Frye}) as 
\begin{flalign}
\langle \Slash{p}^{||}_{T} \rangle _{\rm fluid}
=\!-\!\sum_{i}\!\int\! dp_{T}d\phi_{p} p_{T} \cos \left(\phi_{p}-\phi_{1}\right)
\!\frac{dN_i}{dp_{T} d \phi_{p}}. \label{eqn:fluid-cont}
\end{flalign}
Adding the transverse momentum of the traversing parton pair to this, we obtain $\langle \Slash{p}^{||}_{T} \rangle$
to be compared with the data. 

\emph{Results.}---
A pair of back-to-back partons is supposed to be created 
at $\left(\tau=0, x=x_0, y=0, \eta_{s}=0\right)$
with the common energy, $p^{0}_{a} (\tau=0) = 200\:{\rm GeV}$. 
Until $\tau_{0}$, these partons travel without interacting with the medium.
Then they start to interact with the \textit{expanding} {QGP} fluid at $\tau_0$
and travel
in the opposite direction 
along the $x$ axis.
We can control the jet asymmetric parameter $A_{J}$
by changing the initial position of pair creation: 
When the position of the pair creation 
is off central, namely, $x_0 \neq 0$, 
the amount of energy loss is different between these two partons. 

\begin{figure}[tbp]
 \vspace{2mm}
\begin{tabular}{cc}
\begin{minipage}{0.5\hsize}
\begin{center}
\hspace*{-10.6mm}
\includegraphics[width=57mm]{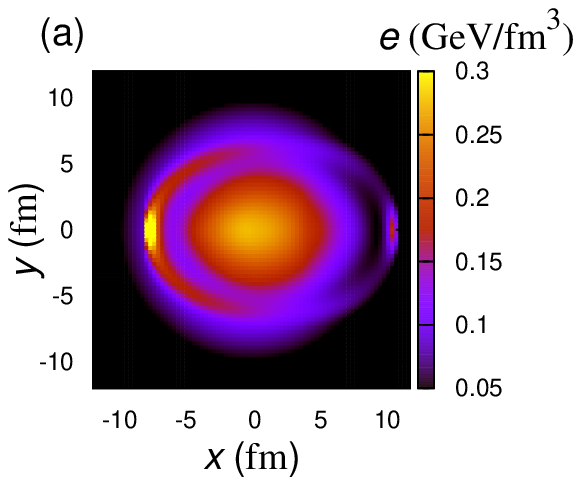}
\end{center}
\end{minipage}
\begin{minipage}{0.5\hsize}
\begin{center}
\hspace*{-11.9mm}
\includegraphics[width=57mm]{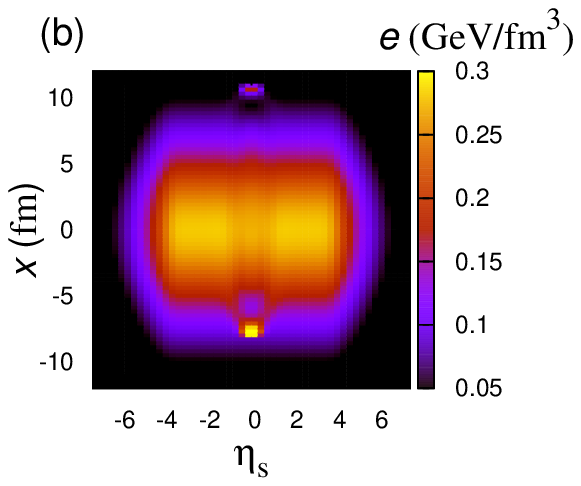}
\end{center}
\end{minipage}
\end{tabular}
 \caption{(Color online) Energy density distribution of the expanding {QGP} fluid 
at $\tau=9.6\:{\rm fm}/c$ (a) in transverse plane at  $\eta_{s}=0$ and (b) in reaction plane at $y=0$.
A pair of energetic partons is created at $\left(\tau=0, x=1.5\:{\rm fm},y=0,\eta_{s}=0 \right)$ and travels in the opposite direction along the $x$-axis at the speed of light. 
}
\label{fig:map}
\end{figure}

Figure \ref{fig:map} shows the energy density distribution 
of the {QGP} fluid at  $\tau=9.6\:{\rm fm}/c$ 
in the transverse plane [Fig. 1(a)] at $\eta_{s}=0$ and in the reaction plane
at $y=0$ [Fig. 1(b)].
Here, the pair of jet is created at $x_0=1.5\:{\rm fm}$
to demonstrate the difference in the amount of energy loss. 
Relatively higher energy density regions exhibits oval structures due to expansion of the QGP,
which are the remnants of Mach cones generated by two energetic partons.
As shown in Fig.~\ref{fig:map} (a), 
these Mach cones are 
distorted by radial flow in the transverse plane.
On the other hand,  
they do not propagate so much apparently in the longitudinal direction
in the reaction plane as shown in Fig.~\ref{fig:map} (b). 
This is simply because 
the coordinate itself expands: 
the Mach cones are spread out 
in the longitudinal direction 
in the Cartesian coordinate. 

We calculate $\langle \Slash{p}^{||}_{T} \rangle$ 
as a function of the dijet asymmetry ratio $A_{J}$. 
The transverse momenta 
of the leading jet $p_{T,1}$ and of the subleading jet 
$p_{T,2}$ 
are estimated 
by summing up 
the transverse momenta of the traversing partons 
at $\tau=9.6\:{\rm fm}/c$ 
and that of particles inside jet cones $\Delta R < 0.5$
obtained through Eq.~(\ref{eqn:fluid-cont}). 
The value of $A_{J}$ changes 
with the position of the pair creation $x_0$. 
\begin{figure}[tbp]
\begin{center}
 \includegraphics[width=60mm]{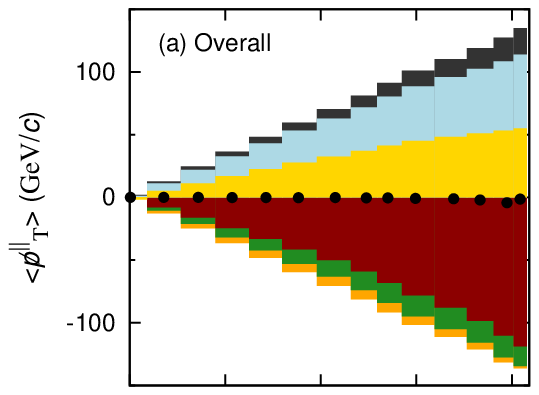}
 \includegraphics[width=60mm]{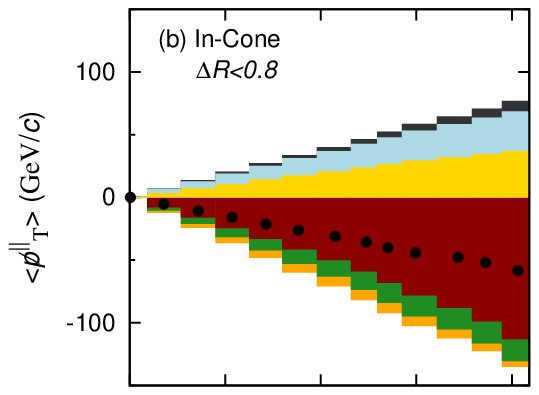}
 \includegraphics[width=60mm]{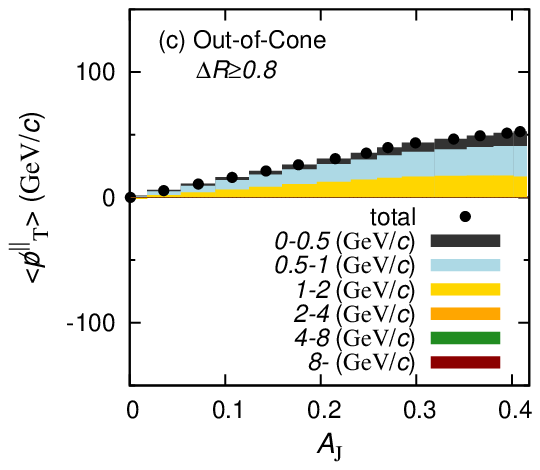}
\end{center}
\caption{(Color online)
$\langle \protect\Slash{p}^{||}_{T} \rangle$ 
as a function of the dijet asymmetry ratio $A_{J}$ 
for (a) the whole region, 
(b) inside the leading and subleading jet cones ($\Delta R < 0.8$), 
and (c) outside 
both of the leading and subleading jet cones  ($\Delta R \geq 0.8$). 
Each band represents the contribution in 
six transverse-momentum ranges: 
$0$-$0.5$, 
$0.5$-$1$, 
$1$-$2$, 
$2$-$4$, 
$4$-$8$ GeV/$c$, 
and $p_{T}>8\:{\rm GeV}/c$. 
The solid circles show 
the total $\langle \protect\Slash{p}^{||}_{T} \rangle$ 
including the contributions mentioned above. 
}
\label{fig:missing}
\end{figure} 
In Fig.~\ref{fig:missing}, 
$\langle \Slash{p}^{||}_{T} \rangle$
as a function of $A_{J}$
 is shown for (a) overall,
 (b) in-cone, and (c) out-of-cone regions of $p_{T}$. 
The solid black circles show 
$\langle \protect\Slash{p}^{||}_{T}\rangle$ 
including the contribution from 
energetic partons 
and 
particles coming from fluids for the whole transverse momentum range. 
As shown in Fig.~\ref{fig:missing} (a), the transverse momentum of the whole system 
is well balanced for any values of $A_{J}$, 
which 
guarantees that 
the energy and momentum 
are conserved 
at very high precision
in the present calculation. 
Each band corresponds to 
the contribution to 
$\langle \protect\Slash{p}^{||}_{T} \rangle$ 
from each transverse-momentum range: 
$0$-$0.5$, 
$0.5$-$1$, 
$1$-$2$, 
$2$-$4$, 
$4$-$8$ GeV/$c$, 
and $p_{T}>8$ GeV/$c$. 
Since the momenta of the energetic partons 
after traveling through the {QGP} fluid are 
still sufficiently large,
they are included in the region of $p_{T}>8\:{\rm GeV}/c$. 
The contribution from the $p_{T}>2\:{\rm GeV}/c$ range 
is negative, \textit{i.e.}, on the leading-jet side. 
This negative contribution is 
balanced 
by the positive contribution 
of the particles with $p_{\rm T}<2\:{\rm GeV}/c$. 
We next analyze 
$\langle \protect\Slash{p}^{||}_{T} \rangle$
inside the jet cone and out of the jet cone separately,
where the two cones with $\Delta R = 0.8$ 
around the leading and subleading jet axes
are considered. 
The contribution of the particles 
inside and outside the cones 
are shown 
in Figs.~\ref{fig:missing}
\ref{fig:missing}(b) 
and \ref{fig:missing}(c), respectively. 
The in-cone 
contribution to 
$\langle \protect\Slash{p}^{||}_{T} \rangle$ 
is negative 
and 
high-$p_{T}$ particles are dominant. 
In the out-of-cone region, 
$\langle \protect\Slash{p}^{||}_{T} \rangle$ 
is positive 
and 
only particles with $p_{T}<2\:{\rm GeV}/c$ 
contribute. 
These out-of-cone low-$p_{T}$ particles 
are originated from 
deposited energy and momentum 
transported 
by the collective flow in the {QGP} fluid. 

\emph{Discussion.}---
The equation of state of massless ideal gas
employed in the present study
might have been too simplified.
However, we find that the basic feature of momentum transport
away from the jet axis due to radial expansion
does not change when we employ the equation of 
state from recent lattice QCD calculations \cite{Borsanyi:2010cj, Borsanyi:2013cga}.
Interestingly, momentum is transported
at larger angles in this realistic equation of state
than in the hardest equation of state employed here
since the softer equation of state
makes the resultant Mach angle  sharper.
This means that we estimate the minimum effect of momentum
transport away from the jet axis.
One would have had to employ the lattice equation of state in a
more quantitative analysis. However, a drawback is
that all the resonances should be considered
at freezeout to see the subtle interplay of momentum balance.
In this Rapid Communication, 
we respect the simple but strict momentum conservation of freezeout processes 
as well as that of hydrodynamic evolution without employing the lattice equation of  state.

It should be noted that the purpose of the present study in this Letter is to demonstrate (and to claim the 
importance of) nonlinear responses 
of the QGP fluid to the jet propagation, 
and that 
we can in principle employ any forms 
of 
source terms, 
which we postpone as future comprehensive studies. 
To determine 
proper source terms for hydrodynamics, 
an investigation of 
energy deposition into the medium 
from in-medium evolution of realisitic jet showers 
and treatment of separation between the hard jet part and the soft fluid part 
are important 
\cite{Neufeld:2009ep, Qin:2009uh, Neufeld:2010tz, Renk:2013pua}.

\emph{Summary.}---
In this Rapid Communication, 
motivated 
by the current experimental situation, 
we studied the collective flow 
in a QGP 
induced by jet particles 
and 
the redistribution dynamics 
of 
the deposited energy and momentum. 
We formulated relativistic hydrodynamic equations 
with source terms 
introduced to
account for deposition of the jets' energy and momenta. 
By solving the hydrodynamic equations 
numerically without linearization 
in fully (3 + 1)-dimensional Milne coordinates, 
we simulated 
dijet asymmetric events in heavy ion collisions. 
In the calculations, 
a new scheme was employed 
to solve the equations  
at very high precision. 
We found that 
jet particles induce Mach cones 
in the medium 
and 
these Mach cones are 
strongly distorted 
by radial flow in the transverse plane, 
but, due to the expanding coordinates, 
not so much apparently in the longitudinal direction 
in the reaction plane. 
We also showed that 
low-$p_{T}$ particles 
are enhanced 
at large angles from the quenched jet axis 
and 
compensated
a large fraction of the dijet momentum imbalance. 
The enhancement 
arises from 
deposited energy and momentum 
transported 
by the collective flow in the {QGP}. 
This fact provides an intimate link between 
theoretical pictures of 
medium response to jet quenching 
and 
the actual phenomenon observed in heavy ion collisions. 
This sheds light on new phenomenological analysis 
to extract the property of the QGP medium 
such as sound velocity and transport coefficients 
by focusing on low-momentum particles 
at large angles from the jet axis. 

\begin{acknowledgements}
The authors thank Y. Hirono, M. Hongo, R. Kurita, and K. Murase 
for useful discussions.
The work of Y.~T. is
supported by a JSPS Research Fellowship for Young Scientists 
and by a grant from the Advanced Leading Graduate Course for Photon Science. 
This work was supported by JSPS KAKENHI Grants No. 
13J02554 (Y.~T.) and No. 25400269 (T.~H.).
\end{acknowledgements}


\end{document}